\def\lsim{\lower -0.3ex \hbox{$<$} \kern -0.75em \lower 0.7ex \hbox{$\sim$}}
\def\gsim{\lower -0.3ex \hbox{$>$} \kern -0.75em \lower 0.7ex \hbox{$\sim$}}
\def\yen{Y \kern -1.077em =}

\documentstyle[12pt]{article}
\title{Quasi-Particles \\in Two-Dimensional Hubbard Model: \\Splitting of
Spectral Weight}
\author{Hidetoshi Fukuyama \\and\\ Masao Ogata$^{\tiny A}$\\ \\
Department of Physics, Faculty of Science, University of Tokyo,\\
7-3-1 Hongo, Bunkyo-ku, Tokyo 113\\
$^{\tiny A}$ Institute of Physics, College of Arts and Sciences, University of
Tokyo,\\
Komaba, Meguro-ku, Tokyo 153}
\date{(Received\hspace{3cm})}
\begin{document}           
\baselineskip=20pt
\maketitle                 

\begin{center}
{\bf Abstract}
\end{center}

It is shown that the energy $ (\varepsilon)$ and momentum $(k)$ dependences of
 the electron self-energy function
 $ \Sigma (k, \varepsilon + i0) \equiv \Sigma^{R}(k, \varepsilon) $ are,
$ {\rm Im} \Sigma^{R} (k, \varepsilon) = -a\varepsilon^{2}|\varepsilon -
\xi_{k}|^{- \gamma (k)}  $
where $ a $ is some constant,
$ \xi_{k} = \varepsilon(k) - \mu,  \varepsilon(k) $
being the band energy, and the critical exponent $ \gamma(k) $,
which depends on the curvature of the Fermi surface at $ k $, satisfies,
$ 0 \leq \gamma(k) \leq 1 $.
This leads to a new type of electron liquid, which is the Fermi liquid in the
limit of
$ \varepsilon, \xi_{k} \rightarrow 0 $ but for $ \xi_{k} \neq 0 $
has a split one-particle spectra as in the Tomonaga-Luttinger liquid.

\noindent
(submitted to J.\ Phys.\ Soc.\ Japan: [Letter])
\newpage

  The nature of low energy electronic excitations in two-dimension is one
of the central theoretical issues in the study of the anomalous metallic
state realized near the Mott insulator.
  Anderson$^{1,2)}$ has proposed that the special feature of the phase volume
in the
scattering processes in
two dimension would lead to a non-Fermi liquid state.
  Anderson argues that the phase shift, $\delta (q, \omega) $, of the forward
scattering
 of two electrons (i.e. the total momentum of two electrons, $q$, approaching
$ 2 \vec k_F, \vec k_F $ being  the Fermi momentum) will stay finite in the
limit of low exciting
energy, $ \omega \rightarrow 0 $, resulting in the singular interaction
leading to the breakdown of the Fermi liquid theory.
  This remarkable suggestion has attracted much interest.$^{3-10)}$
  Especially the many-body perturbational theory based on the $t$-matrix
approximation$^{10,11)}$
has carefully been analyzed.$^{3,4,5)}$ It has been found$^{5)}$ that actually
$\delta(q \rightarrow 2 k_F, \omega \rightarrow 0 )$   is finite if
$q$  and $\omega$  satisfy a particular condition
 i.e.  $(q^2 -4k_F^2)/4m< \omega < q (q-2k_F)/2m $.
  The existence of this singular behavior in a particular scattering processes
has also been
noted in the calculation of Landau $f$-function$^{7)}$ and in the study of
scattering
wave-function in a finite box.$^{7,9)}$
  This can be considered as a partial support to the indication by Anderson.
  It is found, however, that, irrespective of this subtlely together with the
existence
of the bound state,$^{3)}$ ordinary many-body theory for the interaction
processes
confirms that the imaginary part of the electron self-energy  $ \Sigma ^R(k,
\varepsilon ) $
on the Fermi surface    (i.e. $k=k_F$),  is proportional to
 $ \varepsilon ^{2}\ln \varepsilon_F/| \varepsilon|, \varepsilon_F $
 being the Fermi energy$^{3,5)}$ and then the quasi-particle is well defined.
  Here the existence of extra $ \rm ln \varepsilon_{F}/|\varepsilon| $  factor
is to be noted,
which is due to the scattering  processes involving energy ($\omega$)
and momentum ($q$) in the vicinity of the particular values that lead
to a constant phase shift.
  The presence of this factor has first been recognized by Hodges
et al.$^{12)}$ in their study on the
second-order processes, where both forward and backward scattering contribute
to this singular contribution.
  However in the $t$-matrix approximation it is found$^{5)}$
that only the forward scattering can contribute to this singularity, which fact
has
not been noted in the earlier study based on the same approximation$^{13)}$
as has been also stressed recently.$^{6)}$
  This finding implies that the generic singular contributions
to the self-energy function are due only to the forward scattering.

  Another important feature found in ref.\ 5 is that not only the energy
dependence
but also the momentum dependence are singular; i.e.
${ \rm Im} \Sigma ^R(k, \varepsilon ) \propto -\varepsilon ^{2}\ln
\varepsilon_F/| \varepsilon
-  \xi_k | $.
  If this form of  $  {\rm Im} \Sigma ^R(k, \varepsilon ) $  is employed, the
quasi particle
spectral weight,
$ {- \rm Im}G (k, \varepsilon + i0 ) \equiv \rho (k, \varepsilon) $, is split
into two peaks since
$ \rho (k, \varepsilon = \xi_k )=0 $  although the splitting is very small in
this case.
This is consistent with the finding by Castellani et al.,$^{9)}$
who have shown that the diagrams, which lead to the
Tomonaga-Luttinger liquid in one-dimension, do not cause the Tomonaga-Luttinger
liquid in more-than-one dimensions.
As a result, they concluded that in the limit of $( \xi_{k}, \varepsilon)
\rightarrow 0 $,
the splitting   shrinks  faster than $|\xi_k|$  and thus the  Fermi liquid
state is stable.

  In this Letter we will emphasize the splitting of the spectral weight at
finite
$\xi_k$ and $\varepsilon$.
This splitting is given by the particle-particle ladder diagrams which are
canceled out in
one-dimension but have important roles in two dimension.
We will also show that for
a general shape of the two-dimensional Fermi surface the forward scattering
results in a singular contribution to the self-energy function as
$ {\rm Im}\Sigma^R (k, \varepsilon) = -a\varepsilon^{2}|\varepsilon - \xi_k|^{-
\gamma(k)} $
with the critical exponent $ \gamma(k) $  satisfying $ 0 \leq \gamma(k) \leq 1.
$
  This results in a pronounced splitting of $ \rho (k, \varepsilon) $  as a
function of energy,
$ \varepsilon $, although in the limit of  $( \xi_{k}, \varepsilon) \rightarrow
0 $
the  Fermi liquid state is stable.

  The second order contributions to the self energy $\Sigma (k, i\varepsilon_n)
$
    ($\varepsilon_n$ being the thermal energy to be analytically continued,
$i\varepsilon_n\rightarrow\varepsilon+i0$)
is given by the process shown in Fig.\ 1.
  This results in

\begin{equation}
{\rm Im}\Sigma^R (k, \varepsilon )=-U^2 \sum_q \int_0^\varepsilon
 \frac{dx}{\pi} {\rm Im}K(q,x+i0) {\rm Im}G(-k+q, x-\varepsilon-i0),
\label{1}
\end{equation}
where  $G (k, x+i0) $ and $ K (q, x+i0) $ are analytic continuation of the
electron
 Green function and the particle-particle correlation function, respectively.
In the $t$-matrix approximation discussed  above the contributions in the
higher order in $U$
result in the
replacement of $ U $ by $ U_{\rm eff} $, which is constant even in the limit of
 $ U \rightarrow \infty $ on one hand,  and the restriction to $ \vec{q} \simeq
2\vec{k}_F $
(i.e. the forward scattering) for the possible singular contributions on the
other hand.
Here $U_{\rm eff}$ is given by $U_{\rm eff}=U[1+UN(0){\rm ln} k_c/k_F]^{-1}$
where
$N(0)=(4\pi t)^{-1}$, $t$ being the transfer integral in the Hubbard model, is
the density of states
per spin and $k_c$ is the cut-off momentum of the order of the size of the
Brillouin zone.
Hence for $ \varepsilon \rightarrow 0 $, we obtain

\begin{equation}
 {\rm Im} \Sigma ^R(k, \varepsilon ) =
 -\pi U^2_{\rm eff} \int_0^\varepsilon dx x \sum_{k', k''}
\delta (\xi_{k''}) \delta (x-\varepsilon -\xi_{k'}) \delta (x-\xi_{k+k'-k''}).
\label{2}
\end{equation}

Let us focus on some particular point on the Fermi surface, $ \vec k_F, $
nearest to k of our interest as shown in Fig.\ 2 and redefine $ k_x, k_y $
 axes perpendicular and parallel to the tangential direction of the
Fermi surface as in Fig.\ 3,
 with $ \vec k_F $ at the origin. The Fermi surface trajectory will generally
be given by

\begin{equation}
\varepsilon(k) - \varepsilon_F =  vk_x + Ak_y^n =0,
\label{3}
\end{equation}
where $v$ is the Fermi velocity in the normal direction to the Fermi surface,
$A$ some constant and $ n (\ge 2) $ will be an integer (both even and odd) for
 a smooth Fermi surface, which
we assume.

The integration over $ k' $ and  $ k'' $
in eq.\ (2)
is evaluated as follows by use of eq.\ (3),

\begin{eqnarray}
\lefteqn{\sum_{k',k''} \delta (\xi_{k''}) \delta (x-\varepsilon -\xi_{k'})
\delta (x-\xi_{k+k'-k''})} \nonumber \\
&=& {1 \over (2\pi )^4v^2|A|} \int  dk_y'dk_y'' \delta (k''^n_y +(k'_y -
k''_y)^n -k'^n_y - {X \over A} ) \nonumber \\
&=& { C_n \over 8\pi^4v^2n(n-2)|A|} \Biggl| {X \over A}\Biggr| ^{n-2 \over n},
\label{4}
\end{eqnarray}
where $ X = \varepsilon- \xi_{k} $ and $ C_n =1 (2) $ for an even (odd) integer
$ n$ $  (n \geq 2) $. Equation (4) results in

\begin{equation}
{\rm Im} \Sigma^{R}(k, \varepsilon) = -a\varepsilon^{2}|\varepsilon -
\xi_{k}|^{-\gamma(k)},
\label{5}
\end{equation}
with $ \gamma(k) = (n-2)/n $.
Since $ 2\leq n $, we see $ 0 \leq \gamma \leq 1 $.$^{14)}$
  It is seen that if $ n=2 $ which is the
case for most points on the Fermi surface, eq.\ (5) is seen to lead to
 $ \varepsilon ^2 {\rm ln} \varepsilon _F/| \varepsilon - \xi_{k}| $
as it should.
  On the other hand in the region where the Fermi surface is flat
we see $ n \to \infty $   and then  $ \gamma \to 1 $, a similar result to
the case of the marginal Fermi liquid$^{15)}$  as regard the exponent at
$k=k_F$ but with essential difference because of the strong $k$-dependence
in the present case.
  By  use of eq.\ (5), we can evaluate the single
particle spectral weight, $ \rho (k, \varepsilon) $, which has a two-peak
structure
since $ \rho (k, \xi_{k})=0 $ due to $ \Sigma^{R}(k, \xi_{k})=\infty $.
This singularity at $ \varepsilon = \xi_{k} $ in eq.\ (5) will, however,
generally be
removed by the finite life time of electron propagators in Fig.\ 1.
  Although the correct form of this renormalization of the electron propagator
can not easily been assessed, the effects of this renormalization on $ \rho
(k,\varepsilon) $
can be qualitatively estimated by replacing $ \varepsilon - \xi_{k} $
into $ \varepsilon - \xi_{k} + ia_{2}\varepsilon^{2} $ with a some constant
$ a_{2} $ comparable to $a$.
  More explicitly,
$\Sigma^{R}(k, \varepsilon) = -ia\varepsilon^{2}
(\varepsilon - \xi_{k}+ia_2\varepsilon ^2)^{-\gamma}{\rm e}^{i\pi \gamma /2}$
is assumed in order to guarantee that ${\rm Im} \Sigma^{R}(k, \varepsilon)$ is
an even function
of $X \equiv \varepsilon - \xi_{k}$ for a fixed $\varepsilon$ when $a_2=0$.
  An example of $ \rho (k, \varepsilon) $ for a choice of $
a=0.02/\varepsilon_F^{1-\gamma}$
and $ a_{2} =0 $ and $0.02/\varepsilon_F$  for $ \gamma =0.5 $ is shown
in Fig.\ 4.
  (Note that in the limit of $U\rightarrow \infty$,
$a=[4\pi \varepsilon_F{\rm ln}^2(k_c/k_F)]^{-1}$ for $n=2$ and
$a=C_n[4\pi n(n-2)\varepsilon_F^{1-\gamma}{\rm ln}^2(k_c/k_F)]^{-1}$
for $n>2$.)

  As seen there exists a clear double peak similar to the case of the
Tomonaga-Luttinger
liquid.$^{16)}$
  This double peak in the case of Tomonaga-Luttinger liquid reflects the
spin-charge separation.
  In the present case, however, this splitting, which is of order of
$|\xi_k|^{{2 \over 1+\gamma}}$, is reduced faster than the center
of the spectral weight (i.e. at $ \varepsilon = \xi_k $) as $ \xi_{k}
\rightarrow 0 $
in contrast to the case of the Tomonaga-Luttinger liquid, where the splitting
is of the order of $\xi_k$.
  Moreover it is stressed that the exponent, $\gamma$, varies along the Fermi
surface, i.e.
$\gamma (\vec k_F)$.
Strictly speaking, $n=2$  on almost all points on the Fermi surface if it is
smooth.
However, if the Fermi surface is viewed with finite energy  spread, the
effective
value of $n$ will vary  in a wide range.
Hence the theory is self-consistent if the splitting of the spectral weight
evaluated for
a relevant $n$ is larger than this Fermi surface energy spread.
  This fact guarantees the validity  of the
Fermi liquid theory (except the point where $\gamma (\vec k_F)=1$) in the limit
of low
energy excitations as in three dimensional case.
  Hence there exists a {\it crossover} from the Fermi liquid behavior in the
limit of low
excitation energy and the Tomonaga-Luttinger liquid like behavior at finite
energies,
a feature generic to the present two-dimensional electron system.
Note that in one- and three dimension, there is no crossover;
the system is always Tomonaga-Luttinger liquid and Fermi liquid,
respectively, in the energy ranges which we are interested in.
In two-dimension,
the existence of such a crossover in a quasi-particle spectral weight
has not been discussed so far in the studies based on the Fermi liquid theory.
  This result suggests that the temperature dependences of various physical
quantities,
which are averaged over the Fermi surface,
will reflect this non-Fermi liquid behavior at finite temperatures, even though
properties
consistent with the Fermi liquid theory should be seen in the limit of low
temperatures.

  The singular behaviors associated with the forward scattering
found in the present paper is intrinsic to the particular feature
of the phase volume in two-dimension and has nothing to do with the singularity
in the
particle-hole channel, e.g. the nesting.
  This is in contrast to the case for $1\leq d<2$ investigated recently by
Castellani {\it et al.},$^{8)}$
where the particle-hole as well as particle-particle channels play crucial
roles.
  In two-dimension the $t$-matrix approximation is known to be valid
as a low density expansion,$^{10)}$ where the exclusive importance of the
forward scattering has been identified.$^{1,2,5)}$
  As the electron density increases, the importance of the
particle-hole correlations gradually increases.
  Even in this case the $t$-matrix ladder is the key elementary  scattering
processes in the
presence of strong correlation, and the particle-hole correlation should be
considered based
on such particle-particle correlations.
  Hence the present fact that the forward scattering alone can lead to a
non-Fermi liquid
properties at finite energies will be the basis for the further study of the
interaction processes
including the particle-hole correlations, which eventually can lead to various
kinds of
instabilities such as charge density wave, spin density wave and
superconductivity.

\vspace{10mm}

{\bf Acknowledgement}

Authors thank Claudio Castellani, Carlo Di Castro, Walter Metzner and
Phil Stamp for useful discussions in Aspen, where the final part of the
work has been carried out.
This work is  supported by Monbusho Joint Research \lq\lq Theoretical Studies
on Strongly Correlated Electron Systems" (05044037)   and
Grand-in-Aid  for Scientific Research on Priority Areas,
\lq\lq Science of High $T_c$ Superconductivity " (04240103)    and
\lq\lq Novel Electronic States in Molecular Conductors" (06243211)
of Ministry of Education, Science and Culture.

\newpage
\vspace{10mm}
\noindent
{\bf Figure Captions}

\noindent
{\bf Fig.\ 1}  The self-energy correction in the second order of Coulomb
interaction, $U$.

\noindent
{\bf Fig.\ 2}  The schematic representation of a characteristic Fermi surface
(FS)
in two dimensions; $\vec k$ and $\vec k_F$ are the momentum variable of
interest (generally away from FS) and the point on FS nearest to $\vec k$.

\noindent
{\bf Fig.\ 3}  A detailed description of FS near $\vec k$ and $\vec k_F$.

\noindent
{\bf Fig.\ 4}  The energy dependence of the quasi-particle spectral weight,
$\rho (k, \varepsilon )=-{\rm Im}G(k,\varepsilon +i0)$,
for $\xi _k /\varepsilon _F =0.1$ with a choice of
$a=0.02/\varepsilon_F^{1-\gamma}$,
$\gamma =0.5$ and $a_2=0$ (solid line) and $a_2=0.02/\varepsilon_F$ (broken
line).


\begin{thebibliography}{9}
\bibitem{comment}
  The real part of the self-energy correction renormalizes the energy as
$\varepsilon \rightarrow Z\varepsilon$ $(Z>1)$.  This renormalization
naturally comes into eq.\ (5) as $|\varepsilon -\xi_k| \rightarrow
|Z\varepsilon -\xi_k|$.
Since this $Z$-factor is not essential in the following discussions, we put
$Z=1$.
\end{thebibliography}
\end{document}